# Different Perspective on Blue Sky Theory: Theory of Single-Photon Scattering on Bound and Free Electrons

V.V. Semak[1] and M.N. Shneider[2]

[1]Virtual Laser Application Design, LLC

[2]Department of Mechanical and Aerospace Engineering, Princeton University, Princeton, NJ USA

**Abstract**

We present a theory of light scattering consistent with modern physics. We proposed a spatial-temporal model of a photon based on classical model of an atomic oscillator. Using this photon model, we established a criterion for single vs multi-photon irradiation of matter. We demonstrated that the assumption that induced dipole radiation can be of infinitely small power is inconsistent with quantum mechanics. We proposed the energy criteria for scattering of a photon on single and multiple atoms and free electrons. This criterion revealed the limitations and applicability of Rayleigh's and Thomson's models of scattering. According to our theory, light scattering is a threshold process, and the scattering can only take place for those photons with wavelengths shorter than some threshold. Using our model, we computed the loss of energy by a photon with wavelengths longer than the scattering threshold during interaction with bound and free electrons. We showed that a single photon can lose energy in collisionless interaction with atoms, molecules, and free electrons. The new theoretical model predicts the red shift of photon wavelength resulting from such interactions. In particular, it provides an explanation for the blue color of sky and an alternative explanation to cosmological red shift.

1. Introduction

Albert Einstein is attributed with saying "the important thing is not to stop questioning," and "it is a miracle that curiosity survives formal education." Ostensibly, this is in response to a view that many accept what they are taught without exploring the "why". In this progression, opinion becomes fact and evolves to foundational truth. As we advance our understanding, we must constantly check alignment of new ideas and concepts with foundational truths. If they do not align, then either the new concept is incorrect, or the understanding of "established science" must evolve. The true scientist does not fear questioning either. The greatest advances of humanity occur not with incremental knowledge, but with revealing the foundation may be different than what we thought. We suggest such a shift in foundational knowledge that aligns the theory of light scattering with modern quantum mechanics. Our work clearly explains why the Rayleigh and Thomson theoretical models are incompatible with modern physics. It is simple (we address concepts found in standard textbooks), yet not trivial (the implications extend to other areas, such as cosmology).

Driven by practical need, we recently evaluated the Lorentz Oscillator Model (LOM) and discovered multiple concerning shortcomings. First, LOM is unable to provide an accurate qualitative description of the physics of optical phenomena. Second, LOM fails in the quantitative prediction of the optical properties of materials, such as the refractive index and attenuation coefficient as a function of light frequency. Since the Lorentz Oscillator Model is accepted as foundational for the major part of modern physical optics, we were compelled to investigate the reason for such shortcomings. This investigation revealed two seemingly unknown facts [1]. First, it appears Lorentz's model was based on the obsolete paradigm of J.J. Thomson's "plum pudding" atom. Second, according to Lorentz's Nobel Prize acceptance lecture [2], the model that we now call the Lorentz Oscillator Model hinges on an assumption that ether exists and interacts with the electrons in an atom causing friction-like loss of oscillation energy. Our investigation also revealed that since the creation of Lorentz's model (more than a century ago), there has been no research to evaluate whether LOM is consistent with modern physics paradigms. Instead, LOM was accepted without question and a vast body of very complex science built on this unverified foundation. It is as if history were repeating – consider the sophisticated and notably accurate models of celestial dynamics that were built on the foundation of Ptolemy's faulty paradigm. The difference lies in our modern ability to test. Unlike the Ptolemaic scientists, who for the better part of a millennium had no experimental evidences confirming or denying the existence of the crystal spheres, modern physical optics scientists have easy access to multiple verified evidences that the atom is not like J.J. Thomson had imagined, and ether does not exist.

During the past several years we conducted an effort to replace the Lorentz Oscillator Model with a new Atomic Oscillator Model that is consistent with modern physics [3-5]. In this work, one important issue came to our attention – the inadequacy of the term in the electron motion equation that describes the electron oscillation energy loss.

At the beginning of the 20$^{th}$ century, Mandelstam, a contemporary of Lorentz, suggested that the oscillations of the atomic electron, forced by an electro-magnetic wave, are damped due to the re-radiation from the induced time-variable atomic dipole [6]. Such a suggestion produced extensive letter exchange with arguments and discussion between Plank and Mandelstam during that first decade. This exchange was either unnoticed or, most likely, ignored by Lorentz, who continued with the "ether friction" energy dissipation [7]. Surprisingly, the ether friction mechanism that models the damping term as proportional to the first order time derivative of displacement remains in current textbooks as the mechanism for electron energy dissipation. Note that, besides being based on the obsolete paradigm of ether, such interpretation of the damping term contradicts modern electro-dynamics. Indeed, the electron oscillation energy loss due to induced dipole radiation is proportional to the third order derivative of displacement [8]. Previously, we demonstrated that the seemingly insignificant replacement of the first order time derivative with the third order derivative leads to substantial consequences of both physics and mathematics nature [3-5].

Thus, in composing a new Atomic Oscillator Model, one of our foci was to write the re-radiation energy loss term correctly. Note that when formulated in the way proposed by Mandelstam, the induced dipole re-radiation corresponds to the attenuation of the incident light due to scattering. Since the atomic and molecular scattering theory was developed by Rayleigh in the late 19$^{th}$ century, it is logical to question whether this theory was ever revised to agree with quantum mechanics. After all, quantum mechanics was created many decades after Rayleigh proposed his original theory of scattering and more than one decade after he proposed his explanation for the blue color of the sky.

Rayleigh's explanation for why the sky is blue is standard coursework in middle school physics, and numerous popularizers of science give detailed explanations (for example, [9]). All indicators point to its acceptance as foundational knowledge. However, popular acceptance is not the same as proven fact. So, we were driven to ascertain whether this seemingly established blue sky theory was ever reconciled with the foundational assumptions of quantum mechanics. As a result of our investigation, we can say with great degree of certainty that current scattering theories (Rayleigh's and Thomson's) are inconsistent with quantum mechanics. In particular, they are incompatible with the concept of a photon, i.e. the assumption of the quantized character of the emission of electro-magnetic radiation. In the sections below we will demonstrate that Rayleigh's and Thomson's scattering theories are inapplicable for low intensity light, such as in an unfocussed laser beam, sun light, or light emitted by an incandescent source, such as a candle, distant star, or other light sources of similarly low intensity.

It is worth noting the importance of Rayleigh's contribution. He created a theory of scattering on the electrons bound in atoms and molecules by leveraging Maxwell's electro-magnetic theory, which was novel at that time. This significant development laid a foundation for a large part of modern physical optics and led Thomson and Compton to develop their respective theories of scattering on free electrons. However, the use of these theories is limited to academic research that struggles to exploit them in experimental diagnostics that utilize scattering of high intensity coherent light of a laser.

Additionally, we would like to make another note. Imagine irradiation conditions with an average light intensity that is significantly lower than typical focused laser beam intensities. One can suggest that under such conditions the individual photons arrive to the potential "scatterer" without overlapping. Recall what many of us heard anecdotally from our University professors and physicist colleagues that classical electro-magnetic wave formalism corresponds to the multiple superimposing photons. Few realized that this formalism only holds when the intensity of light is above a certain threshold. Even fewer pondered the value of this threshold. Below we will show that this threshold is high. For the irradiances below the threshold, we will propose a new theory of scattering that covers the full range of possible light-matter interaction conditions. This proposal is consistent with modern quantum mechanics, i.e. it includes scattering of a single photon. Obviously, such a general theory of scattering is radically different from what is found in modern optics textbooks. In particular, this new theory provides an accurate description of sunlight scattering, i.e. the scattering of individual photons on atoms and molecules. Thus, this new theory replaces Rayleigh's explanation for why the sky is blue.

There are two additional notes we would like for the reader to consider prior to the description of our approach. First, if one obtains a spectrum function as a pointwise product of the spectrum of radiation of a black body with the temperature of the sun, the spectrum of scattered light as predicted by Rayleigh's theory, and the spectral sensitivity of a human eye; the shape of this function will suggest that the sky should appear green to the human eye, see Figure1 below. Second, our extensive attempt to find any peer reviewed publications reporting the measured spectra of light scattered by atmospheric air yielded zero output. Thus, we conclude that Rayleigh's theory explaining the blue color of the sky was never tested, neither proven nor disproven. While surprising, a plausible explanation could be that Rayleigh's theory rapidly acquired the status of established science rendering no interest in further investigation. After quantum mechanics was invented, the theorists were busy targeting more important problems, leaving the seemingly trivial and "established" blue sky theory behind.

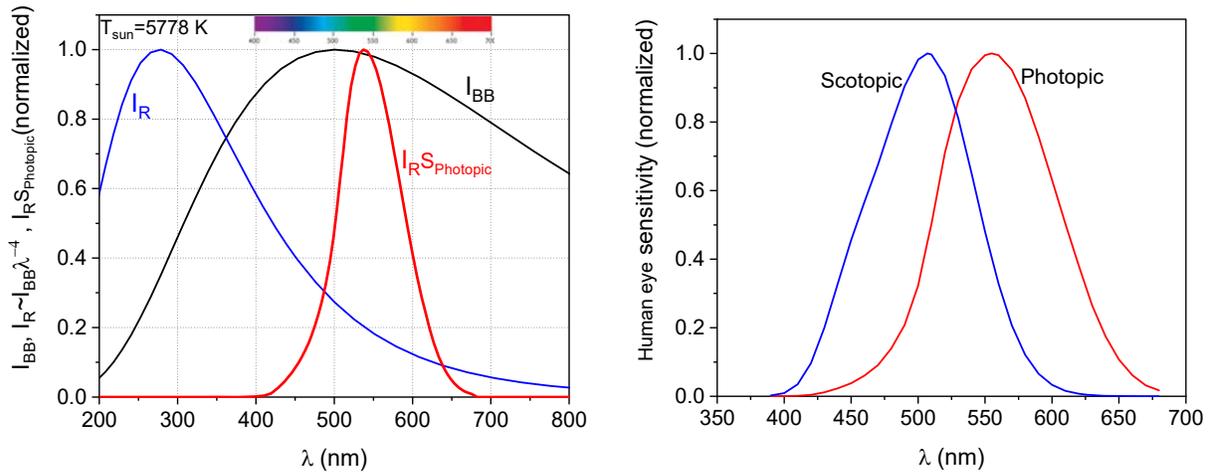

**Figure 1.** a) our calculation of $I_{BB}$ – Planck (black body) spectrum, $I_R$ – Rayleigh ($I_R \propto \lambda^4 I_{BB}$) scattering spectrum, Visible spectrum $\propto I_R S_{photopic}$, b) $S_{Photopic}$ - human eye sensitivity in daylight (Photopic) is calculated based on data from [10].

## 2. Criterion for Single vs Multi-Photon Irradiation

First, let us explain the importance of such a criterion. The theoretical model of light scattering proposed by Rayleigh assumes a continuous function relation between the time-averaged irradiance (light intensity) and the amplitude of the sinusoidal time dependent electric field, i.e. $I_L \propto E_0^2$. In simple terms, Rayleigh assumed that both the light intensity and the corresponding electric field amplitude emitted by an atom could take on values of any magnitude, including infinitesimally small values, continuously down to zero. Extending this assumption further, the scattering is the re-emission of the incident electro-magnetic wave by the induced Hertz dipole, which can produce any intensity of light, including uber-small. Thus, re-phrasing the above, the assumption of Rayleigh's model is that light is always a continuous electro-magnetic wave. Of course, at that time the concept of photons was not yet introduced.

Nowadays, the physicists who grew up with the concepts of the photon and the wave-particle duality of light know that when the average light intensity is low, the individual photons propagate in space without overlapping. If light, indeed, is a flow of photons, then the relationship between the time-averaged irradiance and the electric field amplitude differs from the above relationship formulated for the "continuous" conditions. Consequently, the induced Hertz dipole re-radiation model, which assumes the electric field amplitude value as continuously decreasing down to zero with the decrease of light intensity, would be inapplicable for the description of scattering of a single photon. Note again, the current scattering model remained unchanged since its creation by Rayleigh before the photon concept was coined.

A gap exists between the current understanding of scattering and modern quantum mechanics. We propose to address this gap through developing a theory that includes a description of single-photon irradiation conditions. As a first step, we formulate a criterion that clearly defines multi-photon and single-photon irradiation conditions. In order to develop such a criterion, we need to introduce the characteristic time of the photon duration. This time, denoted as $t_{rad}$, is the time of the radiation emission of a wellezug (wave packet of electro-magnetic radiation) by an emitter, for example, an atomic oscillator described as a Hertz dipole [6,8]. From the equation of an electron oscillator with the damping force that is due to the emission of dipole radiation by the atom [3-6], it follows that the duration of a photon with frequency ω is

$$t_{rad} \sim \left(\frac{\xi}{m_e}\omega^2\right)^{-1}, \qquad (1)$$

where $\xi = \frac{e^2}{6\pi\varepsilon_0 c^3}$. This time of the photon's duration determines the spectral line width customarily called natural broadening, $\delta\nu$:

$$t_{rad} \sim \delta\nu^{-1} = \left(\frac{c}{\lambda^2}\delta\lambda\right)^{-1}, \qquad (2)$$

where $\nu$ and $\lambda$ are the center line frequency and wavelength, respectively, and $\delta\lambda$ is the spectral linewidth.

An important note, below we will use subscript "int" for "interaction" instead of the above subscript "rad". The reason for that is as follows. The time of the radiation emission, $t_{rad}$, is the photon's duration, then, this time is also the time during which the emitted photon interacts with an irradiated atom, molecule or free electron, denoted below as $t_{int}$. Thus, $t_{rad} \equiv t_{int}$.

If a flux of photons irradiates a plane that is normal to the propagation direction, the time between the arrivals of the leading edges of the consecutive photons with center frequency $\omega$ to this plane is

$$t_{ph} = \left(\frac{I_\lambda \, \delta\lambda}{\hbar\omega} S\right)^{-1}, \tag{3}$$

where $I_\lambda$ is the spectral irradiance measured in W/(m² nm), $\delta\lambda$ is the spectral line width, $\hbar$ is the Plank's constant, $\omega$ is the frequency, and $S$ is the area of the atom-photon "collision" cross section.

If $t_{ph} < t_{int}$, the multiple photons overlap while arriving to the observation plane, and in this case $I_L \propto E_0^2$, see Figure 2. The time of arrival of the consecutive photons is inversely proportional to the light irradiance, $I_L = I_\lambda \, \delta\lambda$, and for substantially large irradiance $t_{ph} \ll t_{int}$. In this case, assuming that all the arriving photons have the same phase, the photon flux could be represented as an electro-magnetic wave with sinusoidal time-dependent electric field, $E = E_0 \sin(\omega t + \phi)$. For such conditions, the current theoretical model of scattering, created by Rayleigh, should provide adequate description of scattering of a continuous electro-magnetic wave on an atomic or molecular species. Similarly, for such conditions, a model created by Thomson, should adequately describe the scattering on free electrons.

In the case of low light irradiances, when $t_{ph} > t_{int}$ the single photons arrive to the observation plane without overlapping (see the right wing of the Figure 2). In this case, the atomic or molecular scattering should be described as the scattering of a single photon. Prior to our work presented here, it appears that such a theory for single photon scattering has not been created.

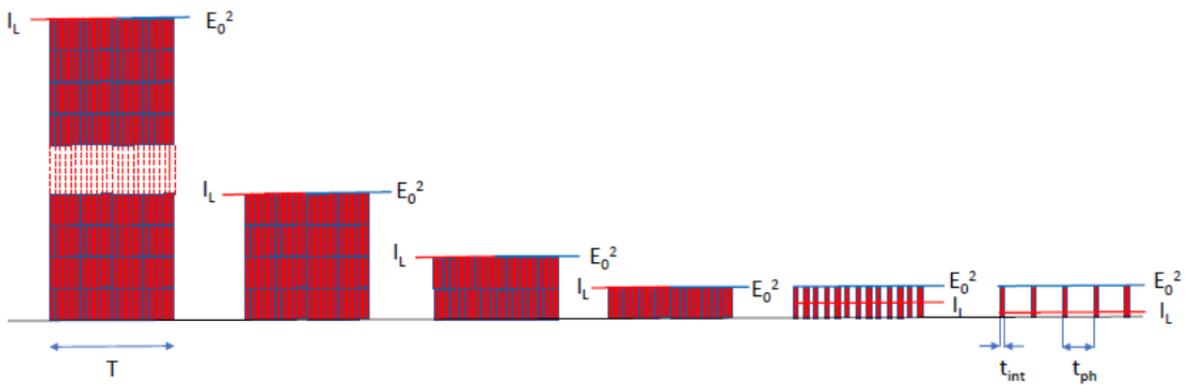

**Figure 2.** Comparison of average light intensity, $I_L$, and squared amplitude of electric field, $E_0'$ for cases of multi-photon (left four) vs single photon (right two) exposures. The photons are represented as the red rectangles with the blue border. The light is assumed to be formatted in pulses of duration $T \gg t_{ph}, t_{int}$, which could be either of finite duration, as in the case of pulsed light, or infinite, as in the case of CW light.

Is there a practical need to develop a theory of single photon (low intensity) light scattering, or does Rayleigh's theory of multi-photon scattering adequately describe most of the conditions of light scattering at the intensities ranging from focused laser light to sunlight? Use of the above equations (2) and (3) provides the answer.

The division of equation (2) by equation (3) yields the ratio of the photon's duration to the time between the arrival of consecutive photons:

$$\frac{t_{int}}{t_{ph}} = \frac{I_\lambda \lambda^2}{\hbar \omega c} S = \frac{I_\lambda \lambda^3}{2\pi \hbar c^2} S. \tag{4}$$

The interaction between an atom and a photon can be treated as a particle collision. Then the interaction cross section is $S = \pi(r_a + r_{ph})^2$, where $r_a$ is the radius of the atom and $r_{ph}$ is the radius of the photon. In the next section we will consider in detail our proposed model of the photon; however, here let us assume that the characteristic diameter of a photon is equal to its wavelength. For the majority of practical needs, the shortest wavelength of an electro-magnetic wave is at least several orders of magnitude larger than the characteristic diameter of an atom. Therefore, for our purpose, the atom – photon collision cross section is $S = \pi\left(\frac{\lambda}{2}\right)^2$, and equation (4) can be re-written as

$$\frac{t_{int}}{t_{ph}} = \frac{I_\lambda \lambda^5}{8\hbar c^2}. \tag{5}$$

Then, from equation (5) it follows that the threshold spectral irradiance of light corresponding to the transition from single photon to multi-photon interaction, in units of W/(m²*nm), is

$$I^t_{\lambda,nm} = \left(\frac{9470}{\lambda_{nm}}\right)^5, \tag{6}$$

where $\lambda_{nm}$ is the wavelength of light in nanometers. If the spectral irradiance of light is lower than the threshold, then the single photon interaction takes place. If the spectral irradiance is larger than the threshold, then the interaction is multi-photon, and if it is significantly larger, then the electro-magnetic wave formalism is applicable for the description of interaction.

Using this criterion, we can ascertain that sunlight scattering in the atmosphere cannot be described by Rayleigh's model because it is suited only for multi-photon interaction. Indeed, the spectral irradiance of sunlight at the wavelength of 500 nm, for example, is approximately $10^6$ times lower than the threshold, see Table 1. Thus, the explanation of why the sky is blue as provided by Rayleigh's model is invalid. At the same time, if a beam of a 0.1 mJ, 100 ns, 10 nm spectral width laser with wavelength of 532 nm is focused into a spot with 10 μm diameter, the laser irradiance in the focal spot will exceed the threshold by more than $10^6$ times. In the latter case, more than one million photons will arrive to an atom at the same time, and then Rayleigh's model is, possibly, applicable. Note that for an unfocussed 10 mm diameter beam of the same laser, the spectral irradiance will be below the single- to multi-photon transition threshold, and again Rayleigh's theory of scattering wouldn't work.

**Table 1.** Threshold spectral irradiance for single to multi-photon scattering transition.

| λ, nm | $I_{\lambda,nm}^{t}$, W/(m²*nm) |
|---|---|
| 100 | 7.62E+09 |
| 300 | 3.13E+07 |
| 500 | 2.44E+06 |
| 700 | 4.53E+05 |
| 1000 | 7.62E+04 |
| 5000 | 2.44E+01 |
| 10000 | 7.62E-01 |

### 3. Photon Model

In order to describe single photon scattering, we need to know the spatial-temporal structure of a photon, i.e. we need a model of photon. The structure of a photon, including its transverse and longitudinal size and the electric field distribution, has received very little attention since its introduction by Lewis in 1926 [11]. Only a few amateur scientists discussed the matter, and that was outside of the peer-reviewed arena [12-14]. In the realm of professional physics, the number of publications that address the structure of a photon is also very small, for example, see [15-19]. The opinion of the theorists [12-16] is that both the longitudinal and transverse sizes of the photon are small, i.e. on the order of one wavelength or much smaller. However, in some very recent publications, experimental physicists [17-19] presented measurement data indicating the length of a photon could be large, possibly exceeding many wavelengths.

Perhaps the lack of interest around the photon model is due to the apparent absence of a practical need to know the photon size and structure. However, in our research of the physics of atomic oscillator interaction with an electro-magnetic wave, we encountered a pressing need for a workable model of a photon. This need stemmed from the attempt to answer a simple question – whether sunlight interaction with air molecules is a single- or multi-photon interaction. What may seem a naïve question is of important practical nature. As we discussed at the beginning of the article, if the interaction of sunlight with an atom or molecule of air is a multi-photon process, then the scattering can be described in terms of re-radiation (scattering) of the electro-magnetic wave by an induced molecular or atomic Hertzian dipole. If, however, the sunlight irradiance corresponds to a single-photon interaction, then the current scattering model is inapplicable and the accepted Rayleigh's explanation of why the sky is blue is incorrect.

In the above section, we introduced equations (1) and (2) as an expression of the time of a photon's interaction with an atom. Let us assume that this time is the characteristic time of the wellezug - a burst of radiation produced by an atomic oscillator. The radiation decay time of an atomic oscillator was first defined by Mandelstam [6] in the early 1900s, and 50 years later it was described in detail in the theoretical physics textbook of Landau and Lifshitz [8]. Since then, it became a trivial concept taught to physics students, as in the case of the famous undergraduate course by Feynman [20]. Curiously, in this lecture Feynman mentioned that quantum mechanical considerations lead to the conclusion that the characteristic time of radiation emission by an atomic oscillator is in the nanosecond range. It is worth noting he did not provide any detailed analysis or discussion, and yet was wondering on the implication that if an atomic oscillator, emitting at ~500 nm wavelength, produces wave trains of nanosecond duration, then the photon with such a wavelength should be several meters long.

Similar to Mandelstam [6] and Feynman [20], we assume that a single photon is produced as a result of the spontaneous emission of a wellezug by an electron in the excited state of an atom. The characteristic time of the radiation decay of the atomic oscillator emission multiplied by the speed of light is equal to the characteristic length of a photon. In the time domain, we suggest that a photon represents a burst of electro-magnetic radiation with a given frequency, ω, and exponentially decreasing amplitude of the electric and magnetic fields with characteristic time of 1/e decay that is $t_{int}$ (Figure 3). This time, $t_{int}$, determines the width of the spectral line of the photon emission, or so called natural line width (see equation (2)). Note that some physical processes, such as elastic collisions, could produce the jumps of the phase of atomic emission if the characteristic (average) time between the collisions, $t_{col}$, is shorter than the atomic oscillator radiation time, $t_{int}$ (see Figure 3). The radiation phase jumps due to such collisions result in the broadening of the atomic emission line. Note, the processes that neither add nor consume the energy of the atomic oscillator emission of the electro-magnetic radiation, such as elastic collisions, would not change the characteristic emission time, $t_{int}$, and therefore, would not affect the length of the photon.

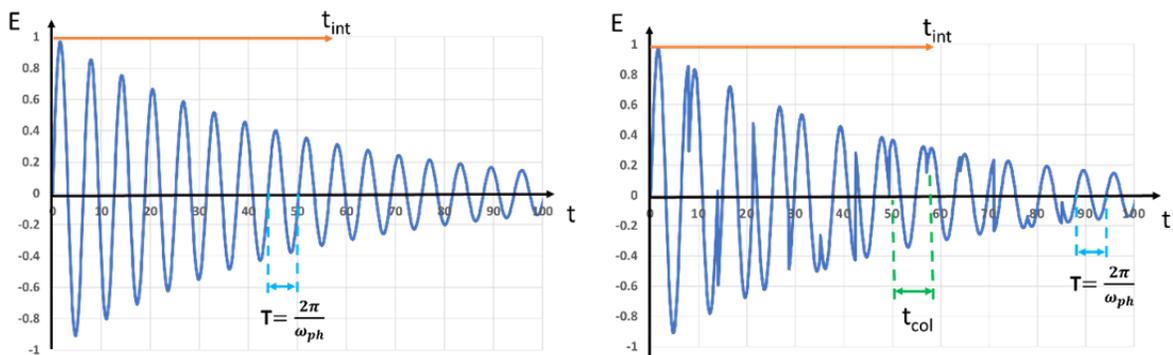

**Figure 3.** Temporal structure of a photon (electric field as a function of time) emitted by an atom without collisions (left) and undergoing collisions during the process of emitting (right).

It is important to note the implications of our discussion so far. If the photon length is assumed to be much shorter than the length defined above, then from the criterion of single-to-multi photon transition (equation (4)), it follows that the multiphoton interaction simply doesn't exist. Indeed, for a photon that is smaller or equal to the wavelength size, in order for the accepted formalism of the electro-magnetic wave to be valid, the irradiance levels would be unreasonably high. To re-iterate, if one rejects the supposition that the photon length is determined by the radiation decay time of an atomic oscillator, and instead one adopts the assumption that the photon length is one wavelength or less [13-17], one must conclude that the description of light as a continuous sinusoidal wave is inadequate for most known conditions. In particular, the assumption of a "very short" photon length leads to the conclusion that both the Rayleigh and Thomson scattering theories are inadequate – this is not only for irradiation, such as in sunlight or unfocussed laser light, as shown above, but also for focused laser light.

In our further consideration of the structure of an individual photon, let us suggest that the field in the photon volume is linearly polarized and the transverse (i.e. normal to k vector) characteristic "diameter" of the photon is approximately $\lambda$. We suggest the transverse distribution of the electric field amplitude has a bell shape (Gaussian, for example) with the radius of approximately $\lambda/2$ on the 1/e level. It is also possible that the characteristic radial size of the photon in the directions that are parallel and perpendicular to the polarization vector are somewhat different. However, we suggest that these characteristic transverse dimensions of a photon are closely approximated by the photon wavelength, at least within the order of magnitude.

Assuming the axisymmetric photon with Gaussian profile of the field amplitude and transverse radius that equals $\lambda/2$ on the 1/e level of field amplitude, and taking into consideration the above discussion of the photon length, the spatial structure of a photon could be as shown in Figure 4.

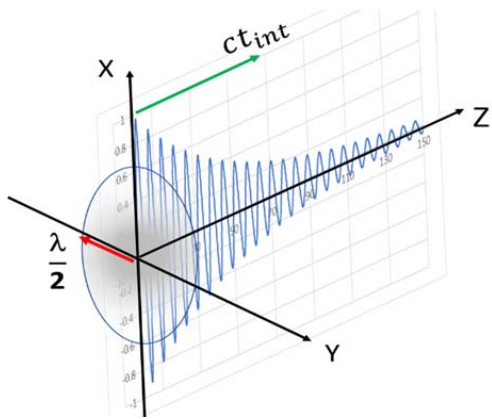

**Figure 4.** Spatial structure of a photon: Gaussian distribution of electric field amplitude as a function of the radial position in the cross sectional plane X-Y with characteristic radius $\lambda/2$, and exponential decrease in longitudinal direction along the Z-axis with characteristic length $ct_{int}$.

Let us now formulate a mathematical model of a photon. We will formulate the set of equations that relate the electric field within the photon volume to the photon energy, starting from the equation for the time dependence of the maximum instantaneous intensity of the electro-magnetic wave:

$$I_0(t) = c\varepsilon_0 \left( E_0 \sin(\omega t + \varphi_{rand}(t, t_{col})) \; e^{-t/t_{int}} \right)^2, \qquad (5)$$

where the commonly used notation for the constants is used, ω is the radiation angular frequency, $\varphi_{rand}(t, t_{col})$ is the phase that is initially zero and then takes random values that are changing at every collision of the radiating atomic oscillator, $t_{col}$ is the characteristic time between the collisions (see the above discussion and Figure 1), and, as above, the duration of the exponentially decreasing amplitude of the electric field in the photon is $t_{int}$ (see equations (1,2)). Then, the energy of a photon is

$$\varepsilon_{ph} = \sigma_{ph} \int_0^\infty I_0(t) dt, \qquad (6)$$

where $\sigma_{ph}$ is the cross section of a photon. If, as we assumed, the radial distribution of the field amplitude is approximately Gaussian with the radius of λ/2 on the 1/e level, then the photon cross section $\sigma_{ph} = \pi \left(\frac{\lambda}{2}\right)^2$. Further, let us assume that the characteristic time between the collisions, those that jolt the phase of an atomic oscillator, is much larger than the time of the period of the radiated photon wave, i.e. $t_{col} \gg \frac{2\pi}{\omega}$. Then, the integral of the radiation intensity in equation (6) can be estimated as follows,

$$\int_0^\infty I_0(t) dt \approx \int_0^\infty c\varepsilon_0 E_0^2 \sin^2(\omega t) e^{-2t/t_{int}} dt \cong \frac{c\varepsilon_0 E_0^2}{2} t_{int}. \qquad (7)$$

Finally, the equation that relates the maximum amplitude of the electric field to the energy of the photon in our model is

$$\varepsilon_{ph} = \sigma_{ph} \frac{c\varepsilon_0 E_0^2}{2} t_{int} = \pi \left(\frac{\lambda}{2}\right)^2 \frac{c\varepsilon_0 E_0^2}{2} t_{int} = \hbar\omega, \qquad (8)$$

where the characteristic time of the emission of radiation by an atomic oscillator, $t_{int}$, is defined above by equation (1).

### 4. Energy criterion for forced dipole re-emission (scattering) of a single photon on an electron bound in a single atom or molecule and on a free electron

As discussed above, the present-day theory of scattering is based on a paradigm where the amplitudes of the electric and magnetic fields in the electro-magnetic wave could take values of any magnitude, including infinitesimally small values, continuously down to zero. This assumption is exactly the same as in the Rayleigh-Jeans theory of blackbody radiation, which contradicts the measurements in the spectral range of short wavelengths and, in particular, predicts the "ultraviolet catastrophe". Of course, as we know now, the assumption of continuously small radiation field amplitudes was replaced by the assumption of the quantum character of radiation. As a consequence of the quantized emission of radiation, there is a lower limit for the amplitudes of the electric and magnetic fields. This lower limit is determined by the energy of the photon and its volume, see Figure 1 for more clarity. If one assumes a "tiny" volume of the photon on the order of $\lambda^3$ or less, as in [11-17], then the minimum amplitude of the electric field would be rather large; and, applying the above single- vs multiphoton criterion, the photon overlap would occur at some unreasonably high light irradiances. At the lower irradiances, the approximation of light-matter interaction as a continuous action of a sinusoidal wave should obviously be wrong, both quantitatively and qualitatively. However, if the volume of the photon is as large as is suggested in our model, then the accepted theories of physical optics have at least some usefulness – they are applicable to the interaction of focused laser light, i.e. for the multi-photon light-atom interaction. In the meantime, the case of "low" intensity, i.e. single-photon interaction, still requires development of an adequate theory. As shown above, the irradiance of the sunlight is "low," and therefore, it is outside of the area of applicability of Rayleigh's scattering theory.

Now, let us present another argument demonstrating that Lorentz's model of the forced atomic oscillator and Rayleigh's and Thomson's scattering theories are inconsistent with the quantum mechanical paradigm of photons. In particular, our argument will demonstrate the impossibility of scattering as the re-emission of the photons. Our argument is based on the supposition that quantum mechanics allows the re-emission of the electro-magnetic wave radiation by the bound or free electrons only if the mechanical energy of the electron oscillations, forced by the photon fields, is no smaller than the energy of the photon.

Let us analyze the conditions under which an induced Hertz dipole can emit. First, recall that when a photon interacts with an atom, the electrons on the outer orbit are displaced from the equilibrium orbit and undergo forced oscillations. This induced dipole is assumed to emit an electro-magnetic wave. It is common textbook knowledge (for example, [8]) that the total power emitted by an oscillating dipole is

$$P_\Sigma = \frac{\omega^4 d_0^2}{12\pi\varepsilon_0 c^3}, \tag{9}$$

where $d_0$ is the atomic dipole moment induced by the incident photon electric field varying with time as a sinusoidal function with frequency, $\omega$. From a theory of an atomic oscillator, either Lorentz's [2] or ours [1,3-5], and assuming negligible loss, the amplitude of the dipole oscillations is

$$d_0 = ea_0 = \frac{e^2 E_0}{m_e(\omega^2 - \omega_0^2)}, \tag{10}$$

where $a_0$ is the amplitude of the electron oscillations and $\omega_0$ is the natural oscillation frequency of the atomic oscillator [1,3-5]. Then, the total emitted power is

$$P_\Sigma = \frac{e^4 E_0^2}{12\pi\varepsilon_0 c^3 m_e^2} \frac{\omega^4}{(\omega^2-\omega_0^2)^2}. \tag{11}$$

Let us assume, as in our model of a photon, that the electric field amplitude decreases with time as $\sim e^{-t/t_{int}}$. If we integrate over time, this gives the energy re-emitted by the atomic oscillator undulating in the field of the photon

$$\varepsilon_{Re} = \frac{e^4 E_0^2}{12\pi\varepsilon_0 c^3 m_e^2} \frac{\omega^4}{(\omega^2-\omega_0^2)^2} t_{int}. \tag{12}$$

In order to satisfy the quantum mechanics fundamental assumptions, this energy re-emitted by the forced atomic oscillator cannot be lesser than the energy of the incident photon that produces the dipole oscillations. If the frequency of the emitted photon is the same as the frequency of the incident photon, as in the case of the elastic scattering, then the energy criterion of whether scattering is possible would be as follows:

$$\varepsilon_{Re} \geq \hbar\omega. \tag{13}$$

In the case of single-photon interaction, the maximum electric field amplitude can be determined from equation (8)

$$E_0^2 = \frac{2\hbar\omega}{c\varepsilon_0 \sigma_{ph} t_{int}}. \tag{14}$$

Substituting equation (14) into equation (12) gives

$$\varepsilon_{Re} = \frac{e^4 \hbar\omega}{6\pi\varepsilon_0^2 c^4 m_e^2 \sigma_{ph}} \frac{\omega^4}{(\omega^2-\omega_0^2)^2}. \tag{15}$$

Dividing both sides of equation (15) by the energy of the incident photon, and taking into account criterion (13), gives the following formulation of the energy criterion for scattering:

$$N = \frac{\varepsilon_{Re}}{\hbar\omega} = \frac{e^4}{6\pi\varepsilon_0^2 c^4 m_e^2 \sigma_{ph}} \frac{\omega^4}{(\omega^2-\omega_0^2)^2} \geq 1. \tag{16}$$

Assuming, as above, that the effective cross section of the photon is

$$\sigma_{ph} = \pi \left(\frac{\lambda}{2}\right)^2 = \frac{\pi^3 c^2}{\omega^2}. \tag{17}$$

Then, the elastic scattering of a photon by an electron bound in a single atom (what is currently referred to as Rayleigh scattering) is possible only for those frequencies of the incident photon that satisfy the inequality

$$N_{bound}^{sa} = \frac{e^4}{6\pi^4 \varepsilon_0^2 c^6 m_e^2} \frac{\omega^6}{(\omega^2-\omega_0^2)^2} = \frac{e^4}{6\pi^4 \varepsilon_0^2 c^6 m_e^2} \frac{\omega_0^2}{(x^2-1)^2} \frac{x^6}{(x^2-1)^2} \geq 1, \tag{18}$$

where $x = \frac{\omega}{\omega_0}$ is the dimensionless ratio of the photon frequency and the natural oscillation frequency of an electron in an atom. Then from equation (18), it is easy to see that the ratio, $N_{bound}^{sa}$, is either equal to or exceeds one for those wavelengths that are shorter than approximately $\lambda_0 = 2\pi c/\omega_0$.

For the case of scattering on a single free electron (currently referred to as Thomson scattering) the natural oscillation frequency, $\omega_0$, is zero and the criterion (18) becomes as follows

$$N_{free}^{se} = \frac{e^4 \omega^2}{6\pi^4 \varepsilon_0^2 c^6 m_e^2} \geq 1. \tag{19}$$

Now let us apply the energy criterion (18) to the light scattering on molecular nitrogen. Previously we demonstrated [1,3-5] that for a hydrogen-like atom in s-state, the natural oscillation frequency of an atomic oscillator is $\omega_0 \cong \sqrt{\frac{2U_0}{m_e r_0^2}}$, where $U_0$ is the ionization potential and $r_0$ is the radius of the electron orbit. As a first approximation, we assume that the above expression for $\omega_0$ can also be used to estimate the natural frequency of oscillations of electrons responsible for photon scattering by simple molecules. Let us approximate the depth of the electronic potential of molecular nitrogen with the value of its ionization potential $U_0 = 15.6$ eV. Another key parameter required for the accurate description of scattering is the value of the effective radius of the electron orbit $r_0$. First, consider that photons with wavelengths less than 190 nm are strongly absorbed in the atmosphere by oxygen molecules. This is well documented and known as the Schumann–Runge absorption bands [21]. Since the scattering of sunlight in the atmosphere undeniably exists, producing the observed blue color of the sky, the scattered photons are weakly absorbed as they propagate through the air. Subsequently, the wavelength of the scattered photons should be greater than 190 nm, i.e. the natural oscillation frequency $\omega_0 \leq 9.92 \cdot 10^{15}$ rad/s. For a nitrogen molecule, such a value of the natural oscillation frequency corresponds to the value of the effective radius of the electron orbit, that is $r_0 \geq 236$ pm. The estimate of the electron orbit of molecular nitrogen ~236 pm is close and exceeds by only 30% the effective radius, computed based on the measured radius of the mean kinetic cross section of the nitrogen molecule ~ 182 pm [22].

Simply put, the Rayleigh theory of scattering is inapplicable in the case of low light intensities that correspond to a single-photon interaction. Regarding the new explanation of the blue color of the sky, the below conclusions follow from our new theoretical model of scattering. First, sunlight intensity is low, corresponding to single-photon interaction across the optical spectrum. Note that in the case of single-photon interaction, there is no reason to consider the combined effects of constructive and destructive interference of incoherent fields of scattered radiation since the wave in the electro-magnetic wellezug of a single photon is coherent. Second, estimation using inequality (18) shows that the scattering of a single photon on a single nitrogen molecule could occur only for the photons with the wavelengths shorter than ~ 200 nm.

## 5. Energy criterion for forced dipole re-emission (scattering) of a single photon on the electrons bound in multiple atoms or molecules and on free plasma electrons

Above we considered one photon interacting with an electron either bound in one atom/molecule or free. This scenario is applicable for interplanetary and interstellar space. Under more common conditions (such as propagation in gases, condensed materials, and high density plasmas), a photon encounters and interacts simultaneously with multiple atoms or molecules and multiple free electrons.

The multiple atoms or molecules exposed to the field of the incident single photon form a common dipole that can scatter the photon. The detailed model of how a dipole is produced from multiple atoms or molecules requires additional theoretical and experimental investigation. As a first approximation, and for the conditions when a scattering dipole is comprised of multiple atomic- or molecular-induced dipoles, the energy criterion for scattering (16) can be re-written as follows

$$N = \frac{\varepsilon_{Re}}{\hbar\omega} = \frac{e^4 N_{id}}{6\pi\varepsilon_0^2 c^4 m_e^2 \sigma_{ph}} \frac{\omega^4}{(\omega^2-\omega_0^2)^2} \geq 1, \qquad (20)$$

where $N_{id}$ is the number of the individual atomic or molecular dipoles. The comprehensive, accurate definition of this number, $N_{id}$, requires additional and detailed investigation. At this point we can propose that the scattering occurs from the number of atoms or molecules contained in the volume of the irradiated material, determined by the photon cross section, $\sigma_{ph}$, and some characteristic length along the k-vector, $l_{ch}$,

$$N_{id} \approx \sigma_{ph} l_{ch} n_a, \qquad (21)$$

where $n_a$ is the species number density.

It is reasonable to suggest that, depending on the particular conditions of the photon-matter interaction, the characteristic length of the photon-matter interaction could be determined by the specific nature of the interaction. The characteristic length, $l_{ch}$, in equation (21) could be the photon wavelength, $\lambda$; the thickness of the skin layer in a conductor, including plasma, $\delta$; the mean free path of gas kinetic collisions in gases or liquids, $l_{fp}$; etc..

For example, if the characteristic length of a photon's interaction with gas is determined by the mean free path length when it is much smaller than the wavelength, $l_{fp} \ll \lambda$, then, recalling the definition of gas kinetic mean free path, $l_{fp} \sim \frac{1}{n_a \sigma_a}$, gives the following approximation for the number of oscillators that interact with the photon

$$N_{id} \sim \sigma_{ph}/\sigma_a, \qquad (22)$$

where $\sigma_a$ is the gas kinetic collisional cross section of air molecules. Then, for the combined multiple atoms dipole scattering of a single photon in a gas with density such that the collision mean free path is much smaller than the wavelength, the energy criterion is as follows

$$N_{bound}^{ma} = \frac{\varepsilon_{Re}}{\hbar\omega} = \frac{e^4}{6\pi\varepsilon_0^2 c^4 m_e^2 \sigma_a} \frac{\omega^4}{(\omega^2-\omega_0^2)^2} = \frac{e^4}{6\pi\varepsilon_0^2 c^4 m_e^2 \sigma_a} \frac{\omega_0^2}{\omega_0^2} \frac{x^4}{(x^2-1)^2} \geq 1. \qquad (23)$$

The estimate calculation shows that the frequency range that satisfies the inequality (23) starts from the lower frequency, as compared to the computed, for a single atom using inequality (16). Thus, our theoretical model predicts that the scattering of a single photon on multiple molecules of nitrogen could occur for wavelengths that are lesser than approximately 350nm.

The scattering of a single photon on the multiple free electrons of the electron-ion plasma could be estimated as follows

$$N_{free}^{me} = \frac{\varepsilon_{Re}}{\hbar\omega} = \frac{e^4}{6\pi\varepsilon_0^2 c^4 m_e^2 \sigma_{el}} \geq 1, \qquad (24)$$

where $\sigma_{el}$ is the cross section of Coulomb electron-ion collision.

### 6. New theory of the sky color

One of the noteworthy deductions that follow from our theoretical model is that single photon scattering is a threshold process. The scattering of a single photon on either one or multiple scatterers is possible only for the photons that have energy exceeding some threshold. This threshold is estimated to be slightly below the value of $\hbar\omega_0$, where $\omega_0$ is the frequency of the natural oscillations of the optical electron in the scatterer (atom or molecule).

As demonstrated above by our new model, the intensiy of sunlight is low, such that the sunlight interaction with air molecules occurs in a single photon regime. Hence, the sunlight scatteing by atmospheric gases is a threshold process that can only take place for those frequencies exceeding some threshold slightly below $\omega_0$, the natural oscillation frequency (see equations (20) and (23)). The estimates for the natural oscillation frequency of molecular nitrogen and molecular oxygen give the upper wavelength threshold for scattering that is estimated to be in the range 190 nm – 350 nm. Note that these two gases comprise 99% of atmospheric gases within the first 100 km of the surface layer. As stated above, the longer wavelengths cannot be scattered in a single-photon interaction since that would contradict the quantum mechanical postulate of the quantized nature of radiation emissions (and the law of energy conservation).

If wavelengths longer than UV light cannot scatter, then the human eye (which is insensitive in the UV spectral range) would see the sky as it appears at night – black with shining stars. So, why is the sky blue? An explanation that does not contradict quantum mechanics is that the scattered UV photons lose part of their energy while propagating in the atmosphere. Below, we compute the residual mechanical energy for both a bound and free electron exposed to the field of a photon. Contrary to popular belief, these computations demonstrate that the energy lost by the photon to induce electron motion is not zero. According to our theoretical model, the photons experience a red shift that increases with the length of the photon path in the atmosphere. This red shift is responsible for the blue color of the sky.

Let us demonstrate how a photon propagating in a medium loses its energy. The equation for forced oscillations of a bound electron in the field of a single photon has the form:

$$\ddot{x} + \omega_0^2 x = \frac{eE_0}{m_e} e^{-\gamma t} \sin(\omega t), \quad (25)$$

with initial conditions $x(0) = 0, \dot{x}(0) = 0$. Note there are several important features of this equation. The equation (25) is missing the radiation loss ("friction") term. Indeed, as we proposed above, and consistent with the assumption of the quantum nature of radiation emission, an electron oscillates without radiation loss if its oscillation energy is less than the energy required to re-emit the photon with energy $\hbar\omega$. In the oscillation equation (25), the forcing electric field is in accordance with our model of a photon, i.e. $E(t) = E_0 e^{-\gamma t}\sin(\omega t)$, and the maximum amplitude can be determined from equation (8). For simplicity, we assume that the photon is produced by the emitting atom without collisional jitter of phase, see Figure 3, left. Finally, the constant $\gamma$ is the reciprocal of the previously defined characteristic time of dipole radiation by an atom, $t_{int}$, given by equation (1).

The analytical solution of equation (25) that satisfies the above initial conditions can be found:

$$x(t) = \frac{eE_0}{m\omega_0} \frac{e^{-\gamma t}\omega_0[(\omega_0^2-\omega^2+\gamma^2)\sin(\omega t)+2\gamma\omega\cos(\omega t)]-\omega[(\omega_0^2-\omega^2+\gamma^2)\sin(\omega_0 t)+2\gamma\omega_0\cos(\omega_0 t)]}{\omega_0^4+2(\gamma^2-\omega^2)\omega_0^2+\omega^4+2\gamma^2\omega^2+\gamma^4}. \quad (26)$$

The equation of motion (26) differentiated over time gives the electron oscillation velocity. When multiplied by the force due to the photon electric field, this gives the instantaneous power acquired from the photon by an oscillating electron

$$P(t) = \dot{x}F = \dot{x}eE_0 e^{-\gamma t}\sin(\omega t). \quad (27)$$

Then, the energy acquired by the electron from the photon is

$$\Delta\varepsilon_{osc} = \int_0^\infty P(t)dt. \quad (28)$$

Note that in practical terms, the infinite upper limit of integration in formula (28) means integration up to time $t \gg 1/\gamma$. The integral in formula (28) can be found analytically; however, the final expresion is ubercumbersome and is of no particular interest due to its triviality. Therfore, for practicality we performed numerical integration of formula (28). Analysis of the results of this numerical intergration shows that the residual energy of the electron oscillations driven by a single photon can be approximated by the value of the kinetic energy of electron oscillation acquired during the first half of the oscillation period (Figure 5):

$$\Delta\varepsilon_{osc} = \frac{1}{2}m_e a_0^2 \omega^2 = \frac{1}{2}m_e \frac{e^2 E_0^2}{m_e^2} \frac{\omega^2}{(\omega^2-\omega_0^2)^2} = \frac{2e^2\hbar\omega}{c\varepsilon_0 m_e \sigma_{ph} t_{int}} \frac{\omega^2}{(\omega^2-\omega_0^2)^2}, \quad (29)$$

where $a_0$ is the maximum amplitude of oscillation. If we set the natural oscillation frequency $\omega_0$ to zero, what remains is the energy transfer from the photon to the free electron for a single-photon interaction with a single electron. The numerical integration of formula (28) demonstrates that for the case of a free electron, i.e. for $\omega_0 = 0$, the residual energy passed from the photon to the electron is exacty equal to the kinetic energy of the electron oscillation acquired during the first half of the oscillation period, and thus, formula (29) is exact.

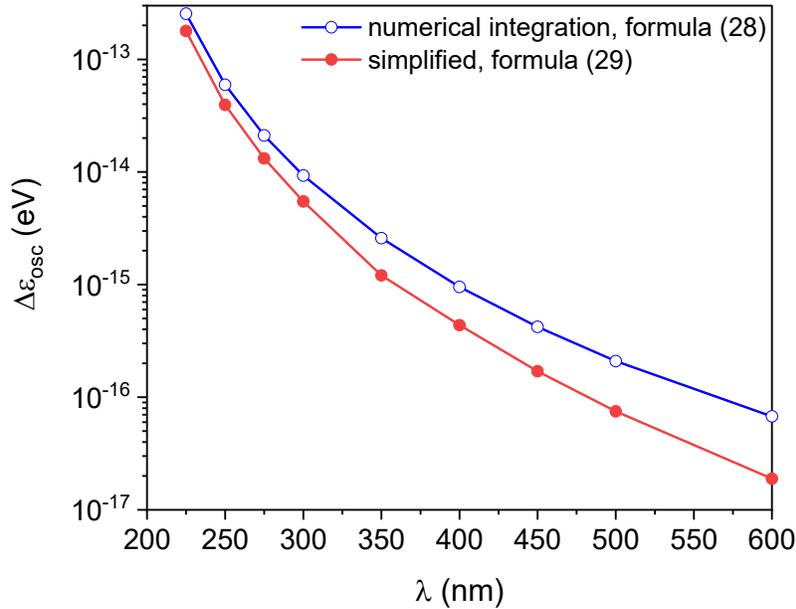

**Figure 5.** Comparison of the computed residual energy of the bound electron driven by the electric field of the incident photon using numerical integration of formula (28) with the approximation as the kinetic energy of the first half of the first oscillation, given by formula (29).

The result that some residual energy is transferred to the electron in a single photon interaction may seem somewhat unexpected. After all, it is commonly belived that an electro-magnetic wave does not transfer energy to an electron in the absence of collisions. Comprehensive theoretical derivations show that if the amplitude of the electric field in the wellezug (photon) were constant, no energy transfer would take place. However, in the wellezug (photon) with decreasing amplitude, some small portion of the energy, $\Delta\varepsilon_{osc}$, is transferred to the electron. This loss of the photon energy leads to a slight "reddening" of the photon that in the case of interaction with a single electron would be immeasurable. However, for a photon traveling long distances through matter, like in the atmosphere, the reddening can become significant.

The results of the calculations performed for the electric field of photons with different wavelengths are given in Table 2. This shows the photon field maximum amplitude $E_0$ computed using equation (8), the characteristic time of the amplitude decrease $\gamma = t_{int}^{-1}$ computed using equation (1), and the photon energy loss $\Delta\varepsilon_{osc}$ in interaction with a single nitrogen molecule computed as a numerical integration of equation (28).

**Table 2.** Photon parameters and energy transferred to a nitrogen molecule

| $\lambda$, nm | $E_0$, V/m | $\gamma$, s$^{-1}$ | $\Delta\varepsilon_{osc}$, eV |
|---|---|---|---|
| 225 | 4799.84 | $4.38 \cdot 10^8$ | $2.54 \cdot 10^{-13}$ |
| 250 | 3688.3 | $3.55 \cdot 10^8$ | $5.94 \cdot 10^{-14}$ |

| 275 | 2906.35 | $2.93 \cdot 10^8$ | $2.11 \cdot 10^{-14}$ |
| 300 | 2338.2 | $2.46 \cdot 10^8$ | $9.30 \cdot 10^{-15}$ |
| 350 | 1509.4 | $1.81 \cdot 10^8$ | $2.58 \cdot 10^{-15}$ |
| 400 | 1139.1 | $1.38 \cdot 10^8$ | $9.53 \cdot 10^{-16}$ |
| 450 | 848.5 | $1.09 \cdot 10^8$ | $4.21 \cdot 10^{-16}$ |
| 500 | 652.01 | $8.87 \cdot 10^7$ | $2.09 \cdot 10^{-16}$ |
| 600 | 413.34 | $6.16 \cdot 10^7$ | $6.75 \cdot 10^{-17}$ |

As the wavelength increases, the fraction of the energy lost by the photon during interaction decreases sharply. Therefore, the photons with wavelengths in the visible and IR ranges experience significantly smaller reddening. In order for a noticeable wavelength shift to occur, the required path length of propagation would exceed the characteristic dimensions of the atmosphere. The largest red shift would occur for those photons with wavelengths in the UV range. This red shift changes the wavelength of the scattered UV photon into the blue part of the visible range. Note that we are considering the red shift of single photons as they propagate in the atmosphere. When a high intensity laser beam propagates in the atmosphere, the redshift may be negligible since the air molecules will be exposed to the coherent multiple photons (see Table 1).

Let us numerically evaluate the magnitude of the red shift as a function of the propagation distance. When a photon interacts with *N* atoms or molecules in its path, the overall decrease of its energy, and hence, the decrease of its frequency, is as follows:

$$\Delta E_{osc} = \hbar d\omega = -\Delta\varepsilon_{osc} N = -\Delta\varepsilon_{osc} \sigma_{ph} n_{air} dl = -\frac{2e^2 \hbar\omega\, n_{air}}{c\varepsilon_0 m_e t_{int}} \frac{\omega^2}{(\omega^2 - \omega_0^2)^2} dl, \qquad (30)$$

where $n_{air}$ is the number density of air. Then, taking into account that the radiation characteristic time (duration) of a photon with frequency ω is given by equation (1), we obtain the equation for the photon frequency decrease (red shift) due to loss of its energy to induce electron oscillations:

$$\frac{(\omega^2 - \omega_0^2)^2}{\omega^5} d\omega = -\frac{2e^2 \xi}{c\varepsilon_0 m_e^2} n_{air} dl, \qquad (31)$$

where, as previously, $\xi = \frac{e^2}{6\pi\varepsilon_0 c^3}$. It is possible to perform exact analytical integration of equation (31); however, for the sake of simplicity let us approximate the threshold frequency of light scattering determined from equations (18) and (23) as smaller than the natural oscillation frequency of the electron, i.e. $\omega_i \ll \omega_0$. Then, the integral of equation (31) provides an estimation of the red-shifted frequency of light after propagating distance *L* in the atmosphere:

$$\omega \approx \frac{\omega_0}{\left(\frac{\omega_0^4}{\omega_i^4} + \frac{8e^2 \xi}{c\varepsilon_0 m_e^2} n_{air} L\right)^{1/4}}. \qquad (32)$$

Substituting the values for the fundamental constants, and assuming that the majority of red shift occurs in the lower levels of atmosphere where the number density of air is $2.55*10^{25}$ m$^{-3}$, gives the frequency change due to the photon energy loss to induce electron oscillations:

$$\omega \approx \frac{\omega_0}{\left(\frac{\omega_0^4}{\omega_i^4}+3.35\ 10^{-3}\ L\right)^{1/4}}, \qquad (33)$$

where the coefficient preceding $L$ is in m$^{-1}$ and the distance, $L$, is in meters.

Converting the frequency to wavelength gives the equation for the wavelength shift

$$\frac{\lambda}{\lambda_i} \approx \left(1 + \frac{8e^2\xi}{c\varepsilon_0 m_e^2} \frac{\lambda_0^4}{\lambda_i^4} n_{air} L\right)^{\frac{1}{4}} \approx 1 + \frac{2e^2\xi}{c\varepsilon_0 m_e^2} \frac{\lambda_0^4}{\lambda_i^4} n_{air} L = 1 + \frac{\lambda_0^4}{\lambda_i^4} 8.37\ 10^{-4} L, \qquad (34)$$

where, similar to equation (33), $L$ is in meters. This equation helps to describe how the sky appears blue. The above considerations of the electron oscillation energy required for photon re-emission demonstrate that light scattering in a nitrogen-oxygen atmosphere can occur for those wavelengths within the range of 200nm – 350nm. The photons with wavelengths less than 190nm are strongly absorbed in the atmosphere by oxygen molecules in the Schumann–Runge absorption bands [21]. Thus, the scattered light has a spectrum distributed in the band 190nm – 350nm. The photons of incident sunlight that have sufficient energy are scattered throughout the upper atmosphere. These scattered photons propagate in all directions while losing their energy as shown above. The human eye, or other spectrally sensitive detector directed away from the sun, would collect these photons scattered from different locations of the upper atmosphere along the line of sight. The longer the distance to the scattering location, the larger the red shift of the wavelength of the scattered photons.

Using equation (34), one can project that the red shift acquired through a ~10km path in the air would convert the 190nm - 350nm wavelength range into the blue colored band, with wavelengths in the range of ~400nm - 500nm. This is a reasonable explanation for the blue color of the sky during the time when the sun is high and irradiating the relatively thin layer of atmospheric air. During sunrise and sunset, when the propagation distances of the scattered sunlight photons are on the order of ~100km, the described wavelength shift converts the 190 nm – 350nm spectral band of the scattered photons into the red colored band. In the meantime, the upper layers of the atmosphere from which the scattered photons propagate into the detector (eye) aperture at distances of ~10km are of the same blue color. This dependence of the color of the sky on the thickness through which the scattered UV photons propagate can be seen during sunrise (see Figure 6).

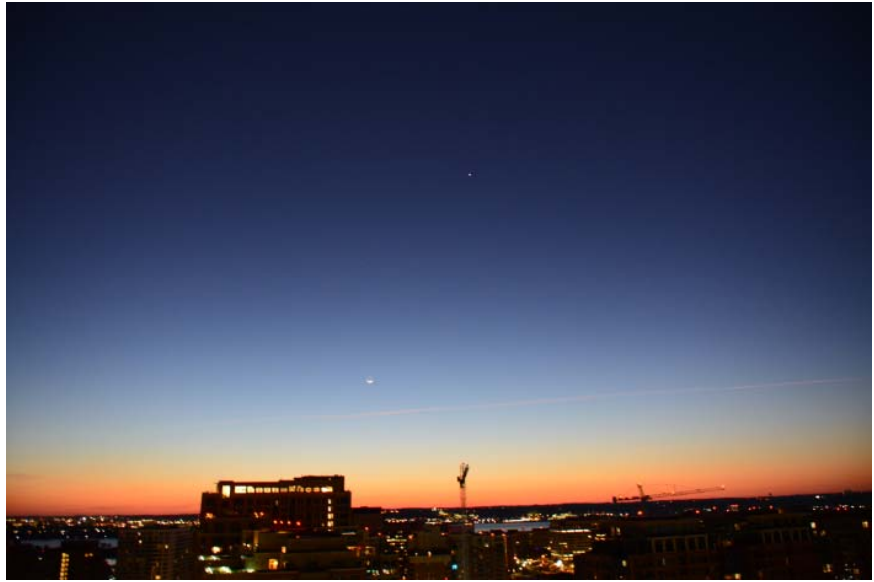

**Figure 6.** Clear sky sunrise over Washington DC, 5:52 am, October 15, 2020; humidity 84%. Courtesy of H.L. Semak.

Of course, the qualitative correlation between the the sky colors in the photograph and the conclusions of our theoretical model cannot serve as a validation of our theoretical model of scattering. A comprehencive comparison of the theoretical spectra with measurement data is required. One would expect that data on the spectra of scattered solar radiation are readily available. However, except for anecdotal accounts, which provide no tangible information on the details of spectrum acquisition and analysis [23], there are no peer reviewed publications in which the measurement results of the spectrum of scattered sunlight are reported. Hopefully, our work presents a reasonable justification for conducting accurate measurements of the scattering spectra of sunlight.

It is possible that a direct validation of our new theory of scattering can be found in the studies of cosmology. It is widely accepted that intergallactic and interstellar space contain free electrons. The application of our proposed theoretical model shows that the photons propagating in intergallactic and interstellar space lose their energy to produce oscillations of the free electrons present there. As the pioneers exploring this area, we call this mechanism of photon energy loss "electronennebeldämpfung". Note that according to our model, the photon loses its energy during interaction with electrons that are either free or bound in atoms or molecules. However, the oscillation energy of a free electron is $10^2 - 10^7$ times larger than the oscillation energy of a bound electron, for example, in hydrogen atoms for the wavelength range from 200nm to 2000nm. Therefore, according to our theory the contribution of intergalactic and interstellar free electron gas to the red shift significantly exceed the contribution of the atomic and molecular species. Using the previously described approach, one can deduce the equation that provides the wavelength at the location of an observer for the case of interction with free electron gas

$$\lambda = \lambda_i \exp\left(\frac{2e^2\xi}{c\varepsilon_0 m_e^2} n_e L\right), \tag{35}$$

where $\lambda_i$ is the wavelength of the emitted photon, $n_e$ is the average density of free electrons in the universe, and $L$ is the distance from the emitter to the observer. Since the values of the term in the exponent are much smaller than one, equation (35) can be approximated using the first two terms of the Taylor series as follows:

$$\lambda = \lambda_i \left(1 + \frac{2e^2\xi}{c\varepsilon_0 m_e^2} n_e L\right). \tag{36}$$

The variable second term in equation (36) is linearly dependent on the distance from the observer to the emitter. Thus, equation (36) demonstrates that the red shift, $\Delta\lambda/\lambda_i$, of the spectral line centered at $\lambda_i$ would be observed as a liner function of the distance from the observer to the emitter star, *L*, due to the photon loss of energy in the interactions with free electrons floating in the universe. Currently in cosmology, the observed red shift of spectra of light emitted by distant stars is asumed to be due to the Doppler effect

$$\lambda = \lambda_i \left(1 + \frac{u}{c}\right) = \lambda_i \left(1 + \frac{H_0 L}{c}\right), \tag{37}$$

where $H_0$ is the Hubble's constant and u is the speed of expansion of the Universe. Equating the formulae (36) and (37) leads to the following formula for the Hubble's constant

$$H_0 = \frac{2e^2\xi}{\varepsilon_0 m_e^2} n_e. \tag{38}$$

It is estimated that the average density of free electrons in the universe ranges from $10^2$ m$^{-3}$ in intergallactic space to $10^4$ m$^{-3}$ in inersterllar space [24]. A typical galaxy size is ~30,000 light years and typical distance between galaxies is ~ $10^6$ light years [25]. Using these distances to calculate the proportional contribution, we can estimate the average free electron density along the path of a light beam through the universe to be $n_e \sim 10^3 \, m^{-3}$. Substituting this number in equation (38) renders a value for $H_0$, ~ 68 km/s/Mpc – remarkably close to the currently accepted range for the values of the Hubble constant, 69.8 +/-1.9 (km/s)/Mpc - 73.3+5.3/−5.0 (km/s)/Mpc.

7. **Broad Implications and Possible Future Research**

Although current scattering theories (Rayleigh's and Thomson's) provide a reasonable approximation for the conditions of multiple coherent photons' irradiation of matter, i.e. conditions corresponding to the focused beam of a laser with high power or pulse energy and narrow linewidth, these theories are incompatible with the postulates of quantum mechanics and cannot be used for single-photon scattering. In view of our research, more thorough research work is required to further develop and validate our theory of scattering that is fully consistent with quantum mechanics at any irradiation intensity.

Our model of a photon leads to a conclusion that atmospheric gas-kinetics contribute to the color of the sky. Imagine that a cluster of the molecules scattering a photon moves out of the photon path. Would the photon be "cut" by such motion of if the scatterers? More theoretical research and experiments are needed to study the effect of gas particle motion on the photon propagation in media. Such research would also target the verification of our theoretical model of a photon. We believe that such experiments are within modern experimental capabilites.

Another possible research avenue that is inspired by this work is the interaction of multiple incoherent photons with matter. In the course of our work we discovered that this is an unexplored subject. While a literature search produces multitudes of works discussing "incoherent scattering," closer examination shows that the topic of these papers is actually the "scattering of coherent light on randomly distributed scatterers." Such problem would have been better stated as the "incoherent scattering of a coherent electro-magnetic wave." The problems of "incoherent scattering of a coherent wave" and "scattering of an incoherent wave" are different problems. The latter has yet to be considered.

## 8. Conclusions

A theoretical examination, in terms of light irradiance and photon energy, showed that under the most common conditions of optical irradiation (such as sunlight, incandescent sources, and unfocussed laser beams) the interaction corresponds to a single photon interaction with an atom, molecule or free electron. All modern theoretical models of light-matter interaction, including the scattering models, assume an atom interacting with a sinusoidal electro-magnetic wave. Anecdotally (as we have heard from our teachers), the continuous sinusoidal electro-magnetic wave corresponds to the irradiation with simultaneously incident multiple photons. In order to fill the gap in the physics of light-matter interaction, we composed a theoretical model of single photon interaction based on our previously developed Atomic Oscillator Model [1].

For the successful application of our Atomic Oscillator Model, a spatial-temporal structure of a photon is needed. Here, we proposed a theoretical model of a photon structure that stems from the description of the emission of an electro-magnetic wellezug by an atom.

We concluded in this work that an atom can emit a wellezug of electro-magnetic radiation (a photon) only if the electron oscillation energy is equal to, or exceeds, the energy of the photon, $\hbar\omega$. Therefore, we concluded that the single photon scattering, such as sunlight scattering, is a threshold process. The scattering occurs only for sufficiently energetic photons. Our theory shows that in atmospheric air (molecular nitrogen), wavelengths longer than ~ 200 nm – 350 nm cannot be scattered since the photons are not energetic enough to produce electron oscillation with the energy required for re-emission.

We composed a theoretical model for photon energy losses during propagation in an atomic/molecular matter and in the free electron gas. Our theoretical model showed that when the photon energy is insufficient to produce scattering via photon re-emission, the photon losses its energy without scattering. We applied this theory to two important scenarios. First, we showed that the photons emitted as a result of scattering by molecular nitrogen lose their energy while propagating in air (note, the propagation of the scattered photons occurs without secondary scattering). Due to this loss, the scattered photons with wavelengths less than 200 nm – 350 nm experience red shift – they become blue. This explains the color of the sky. Second, we showed that the interaction of photons with free electrons results in photon energy loss. Related to this energy loss, the red shifted wavelength is proportional to the original wavelength and exponentially dependent on the product of the length of propagation and electron density. The argument of the exponent function is substantially smaller than one, and therefore, the dependence can be approximated as a function linearly dependent on the propagation distance. This presents an alternative explanation for the Hubble red shift, implying a stationary universe. Note, the deduced exponential dependence of the wavelength on the propagation distance noticeably deviates from the linear approximation at long distances. This provides a different explanation for the observations currently attributed to the accelerated expansion of the outer regions of the Universe.

**Data accessibility.** All data are provided in full in the paper

**Authors' contributions.** The authors contributed equally to this paper

**Competing interests.** We declare we have no competing interests

**Acknowledgements**

The authors would like to thank Heather Semak for support and assistance in preparation of this manuscript.